\documentclass[conference]{IEEEtran}

\usepackage{amsmath,amssymb,amsfonts}
\usepackage{algorithm}
\usepackage{algorithmic}
\usepackage{cite}
\usepackage{url}
\usepackage{amsthm}
\usepackage{soul}
\usepackage{xcolor}
\usepackage{graphicx}
\usepackage{subcaption} % for subfigures
\usepackage[
    top=0.75in,
    bottom=1.1in,
    left=0.70in,
    right=0.70in
]{geometry}

\title{Beyond RSRP Coverage: UAV Serviceability in Interference-Limited Urban 5G NR Networks}
\IEEEoverridecommandlockouts
\author{
\IEEEauthorblockN{Md Sharif Hossen, Vijay K. Shah and Ismail Guvenc}
\IEEEauthorblockA{Department of Electrical and Computer Engineering\\
North Carolina State University, Raleigh, NC, USA\\
Email: \{mhossen, vijay.shah, iguvenc\}@ncsu.edu
}
\thanks{This work is supported in part by the NASA ULI award 80NSSC25M7102 and INL LDRD award DE-AC07-05ID14517. Corresponding author: Ismail Guvenc (email: iguvenc@ncsu.edu).}
}
\begin{document}
\vspace{-5in}
\maketitle
\begin{abstract}
Reference signal received power (RSRP) is commonly used to assess 5G cellular coverage, but it may substantially overestimate the connectivity available to cellular-connected unmanned aerial vehicles (UAVs) under strong inter-cell interference (ICI). As a UAV ascends, reduced blockage can strengthen the serving signal, while increased line-of-sight visibility to neighboring sectors can intensify ICI. This paper develops a trajectory-aware, 3GPP-based system-level framework for evaluating UAV radio serviceability during a three-dimensional flight in an urban 5G New Radio network. The framework jointly evaluates RSRP, reference signal received quality (RSRQ), and signal-to-interference-plus-noise ratio (SINR) in a 19-site, 57-sector deployment. We introduce the concept of \emph{coverage-serviceability gap} to quantify the difference between RSRP availability and joint service availability. Under RSRP, RSRQ, and SINR thresholds of $-100$ dBm, $-18$ dB, and $0$ dB, respectively, an inter-site distance (ISD) of 500 m provides 99.9\% RSRP availability but only 13.6\% joint service availability, yielding a coverage-serviceability gap of 86.3 percentage points. With RSRQ and SINR thresholds of $-22$ dB and $-6$ dB, respectively, joint availability increases to 55.0\% and 63.1\% for ISDs of 500 m and 1000 m, respectively. These results demonstrate that RSRP-based coverage substantially overestimates joint UAV serviceability under both evaluation configurations, with SINR remaining the limiting metric.
%Under evaluation thresholds of $-100$~dBm for RSRP, $-20$~dB for RSRQ, and $0$~dB for SINR, an inter-site distance (ISD) of 500~m provides 99.9\% RSRP availability but only 13.6\% joint service availability, yielding a gap of 86.3 percentage points. Increasing the ISD to 1000 m improves mean SINR and joint availability from 13.6\% to 18.4\%, while SINR remains the dominant constraint. Results averaged over three independent simulation realizations show that RSRP-based aerial coverage does not necessarily translate into mission-level radio serviceability, motivating trajectory-aware, multi-key performance metric (KPM) aerial radio planning. %Increasing the ISD to 1000~m improves SINR availability from 13.6\% to 18.4\%, while mean joint availability also increases to 18.4\% because SINR remains the dominant constraint. Results averaged over three independent simulation realizations show that RSRP-based aerial coverage does not necessarily translate into mission-level radio serviceability and motivate trajectory-aware, multi-key performance metric radio planning for reliable UAV connectivity.
\end{abstract}

\section{Introduction}
\label{sec:introduction}
Cellular-connected unmanned aerial vehicles (UAVs) are increasingly considered for surveillance, emergency response, infrastructure inspection, data collection, and mission-critical command-and-control (C2) applications~\cite{mozaffari2019tutorial,fotouhi2019survey}. Using existing cellular infrastructure can provide wider coverage, lower deployment cost, and more direct integration with network management than dedicated point-to-point UAV communication systems. However, terrestrial cellular networks are designed primarily for ground user equipment (GUE), and their performance characteristics can change substantially when a user moves into three-dimensional (3D) aerial space.

A key challenge originates from the downtilted sector antennas employed by terrestrial base stations (BSs). While their main lobes are directed toward terrestrial users, aerial users are frequently served through antenna sidelobes. At low altitude, buildings and other obstructions can weaken the serving link. As altitude increases, reduced blockage and higher line-of-sight (LOS) probability can strengthen the received signal from the serving sector. The increase in visibility, however, exposes the UAV to a larger number of neighboring sectors operating on the same carrier, which can substantially increase inter-cell interference (ICI)~\cite{Maeng11003933}. Consequently, strong received power does not necessarily indicate that a UAV has sufficient signal quality for reliable communication.

Cellular coverage is commonly characterized using reference signal received power (RSRP). This approach is reasonable for identifying locations where a serving-cell signal can be detected, but it does not directly account for aggregate interference. Reference signal received quality (RSRQ) and signal-to-interference-plus-noise ratio (SINR) provide complementary information because they reflect the relationship between the desired signal, interference, and receiver noise. For aerial users, an RSRP-based coverage map may therefore identify a large portion of the airspace as covered even when RSRQ or SINR is insufficient for usable service. This discrepancy is particularly important for UAV missions because a C2 link must remain available throughout a trajectory rather than only at isolated locations.

Prior studies have established the altitude-dependent propagation and interference characteristics of cellular-connected UAVs through analytical models, simulations, and measurements~\cite{Maeng11003933,Jie10587227}. Standardization studies have also provided channel and deployment models for aerial cellular evaluation~\cite{3gpp38901,3gpp36777}. Existing work has examined UAV coverage probability, antenna sidelobes, BS cooperation, mobility, and interference mitigation~\cite{Chowdhury2021ReliableUAV,wang2022bscoop,hossen_booster-cell2026}. These studies provide an essential understanding of aerial cellular links, but the performance is often reported through individual key performance metrics (KPMs), coverage probabilities, or evaluations at a set of fixed aerial locations. Comparatively less attention has been given to quantifying whether multiple radio requirements are satisfied simultaneously throughout a 3D UAV mission. As a result, the difference between nominal RSRP coverage and mission-level radio serviceability remains insufficiently characterized.
\begin{figure*}[!t]
    \centering

    \subfloat[Urban cellular deployment and horizontal UAV route.]{
        \includegraphics[width=0.52\textwidth]
        {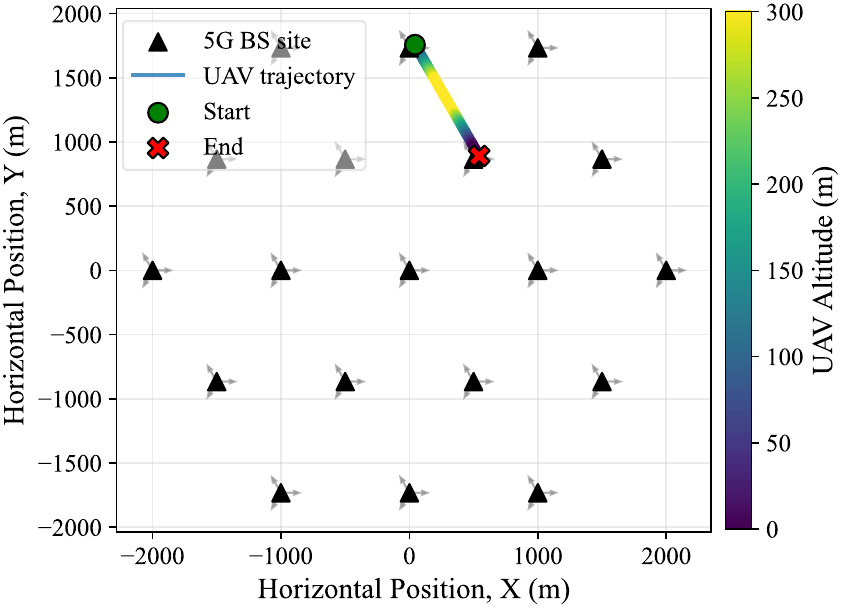}
        \label{fig:deployment_2d}
    }
    \hfill
    \subfloat[3D UAV flight trajectory.]{
        \includegraphics[width=0.42\textwidth]
        {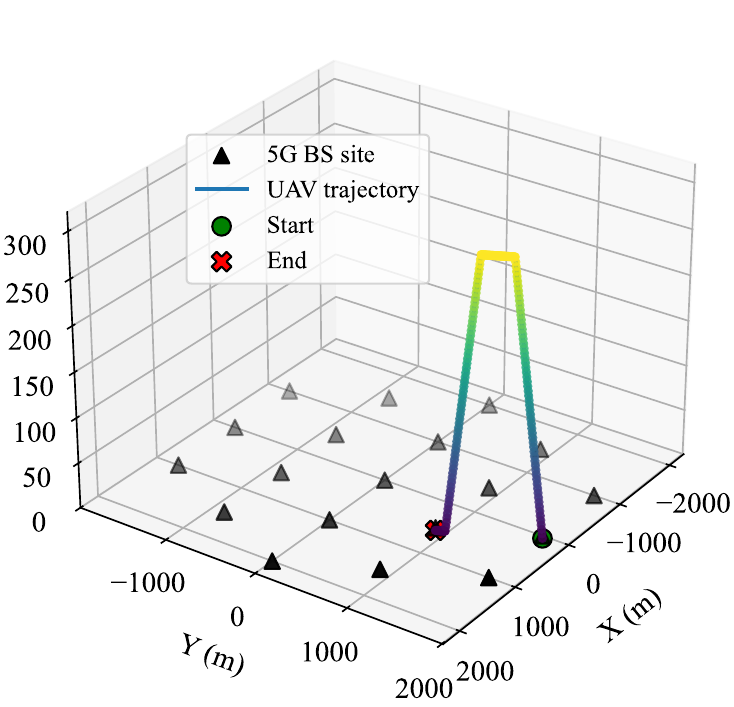}
        \label{fig:trajectory_3d}
    }

    \caption{Simulated urban cellular deployment and UAV mission for a representative ISD of 1000~m. The network contains 19 sites with three sectors per site. The arrows indicate the sector boresight directions, while the trajectory color represents UAV altitude. The UAV moves horizontally while following a climb-cruise-descent altitude profile. }
    \label{fig:network_and_trajectory}
    \vspace{-.2in}
\end{figure*}
~
~
This paper addresses this issue through a trajectory-aware, 3GPP-based system-level analysis of cellular-connected UAV serviceability in an urban 5G New Radio (NR) network. We consider a 19-site, 57-sector urban macrocell (UMa) deployment with co-channel interference, directional sector antennas, probabilistic LOS and non-LOS (NLOS) propagation, and multiple inter-site distances (ISDs). The UAV follows a 3D trajectory consisting of climb, horizontal flight, and descent phases. Along the trajectory, the serving sector evolves according to an RSRP-based A3 handover procedure. The corresponding serving-cell RSRP, RSRQ, and SINR are then evaluated.

Rather than treating these KPMs independently, we define radio serviceability as the simultaneous satisfaction of selected RSRP, RSRQ, and SINR requirements. We further introduce the~\emph{coverage-serviceability gap}, denoted by $\Delta_{\mathrm{CS}}$, as the difference between RSRP availability and joint multi-KPM availability. This metric quantifies how much an RSRP-only assessment overestimates the portion of a UAV mission that meets both signal strength and signal quality requirements. The per-KPM availability comparison further identifies the dominant serviceability bottleneck and how it changes with network density.

The main contributions of this work are as follows:
\begin{itemize}
    \item We develop a trajectory-aware, 3GPP-based system-level framework for evaluating cellular-connected UAV performance throughout a 3D flight in an urban multi-site 5G NR network.

    \item We formulate joint radio serviceability using RSRP, RSRQ, and SINR requirements and introduce the coverage-serviceability gap to quantify the discrepancy between nominal RSRP coverage and mission-level multi-KPM availability.
    \item We quantify the effects of ISD and selected signal-quality thresholds on joint UAV service availability. The results show that increasing ISD can improve SINR by reducing aggregate ICI, while joint service availability remains limited by SINR under both evaluation configurations.
    %\item We quantify the effect of ISD on signal-strength availability, signal-quality availability, and the dominant serviceability bottleneck. The results show that increasing ISD can improve SINR by reducing aggregate ICI, while joint serviceability may remain limited by SINR.
\end{itemize}
% The numerical results reveal a substantial mismatch between RSRP-based coverage and joint radio-condition availability. Under the selected evaluation thresholds, the 500-m-ISD deployment provides 99.4\% RSRP availability along the UAV trajectory but only 3.3\% joint availability, corresponding to a coverage-serviceability gap of 96.1 percentage points. Increasing the ISD to 1000 m improves SINR availability from 15.0\% to 27.8%, while joint availability increases only to 3.9% because RSRQ remains the limiting KPI. These results demonstrate that strong serving-cell RSRP alone does not adequately characterize interference-limited aerial radio conditions and motivate trajectory-aware, multi-KPI radio planning for urban UAV operations.

The remainder of this paper is organized as follows. Section~\ref{sec:system_model} presents the urban cellular deployment, UAV trajectory, antenna configuration, and propagation model. Section~\ref{sec:serviceability} defines the radio KPMs, joint serviceability, and coverage-serviceability gap. Section~\ref{sec:results} presents the numerical results and discusses the effects of altitude and ISD. Finally, Section~\ref{sec:conclusion} concludes the paper.

\section{Urban Network and UAV Mission Model}
\label{sec:system_model}

This section presents the urban cellular deployment, 3D UAV mission, sector antenna model, and large-scale propagation assumptions. The model is based on the UMa geometry in 3GPP TR~38.901 and the aerial-user considerations in 3GPP TR~36.777~\cite{3gpp38901,3gpp36777}.

\subsection{Multi-Site Urban Deployment}
\label{subsec:network}
We consider a cellular network containing $N_{\mathrm{BS}}=19$ BS sites. Each site contains three sectors separated by $120^{\circ}$ in azimuth, resulting in the sector set $\mathcal{S}$, where $|\mathcal{S}|=57$. All sectors operate on the same carrier frequency and bandwidth and, therefore, generate co-channel interference. The ISD, denoted by $D_{\mathrm{ISD}}$, is varied to examine the effect of network density on aerial signal strength and quality. Let the horizontal position of sector $s\in\mathcal{S}$ be $\mathbf{b}_s =
    \begin{bmatrix}
        x_s\;\;y_s
    \end{bmatrix}^{\mathsf T},$
where $x_s$ and $y_s$ denote the horizontal cartesian
coordinates of sector $s$. The three sectors belonging to the same BS site share the same horizontal coordinates but have distinct antenna boresight directions $
    \psi_s \in \left\{0^\circ,120^\circ,240^\circ\right\}.$
All sector antennas are installed at the BS height $h_b$, where $h_b$ denotes the height of the sector antenna above ground level. The evaluation trajectory is selected within the interior of the 19-site deployment so that the UAV remains surrounded by multiple serving and interfering sectors. Fig.~\ref{fig:network_and_trajectory}(a) shows the horizontal UAV route within a representative 19-site deployment with an ISD of 1000~m. The interior route keeps the UAV surrounded by candidate serving and interfering sectors, while the same geometry is scaled for each evaluated ISD.

\subsection{Three-Dimensional UAV Mission}
\label{subsec:trajectory}
The UAV follows the 3D trajectory illustrated in
Fig.~\ref{fig:network_and_trajectory}(b). Its horizontal position changes along the selected route, while its altitude follows a climb-cruise-descent profile. The UAV first ascends toward
the cruise altitude while advancing along the horizontal route. It then maintains an altitude of 300~m for a portion of the flight before descending while continuing toward the destination. The trajectory
is sampled every $\Delta t=0.1$~s, and the horizontal UAV speed is $v_{\mathrm{UAV}}=5$~m/s. The complete trajectory includes takeoff and landing samples near ground level. For the radio-serviceability analysis, only samples with UAV altitudes between 5 and 300~m are retained. This trajectory construction enables the evaluation to capture changes in propagation, antenna alignment, interference exposure, and serving-sector
association throughout a continuous 3D flight. The UAV trajectory is given by:
\begin{equation}
    \mathbf{u}[n]
    =
    \begin{bmatrix}
        x[n] & y[n] & h[n]
    \end{bmatrix}^{\mathsf T},
    \qquad
    n\in\{0,\ldots,N_{\mathrm{T}}-1\},
\end{equation} 
where $x[n]$ and $y[n]$ denote the horizontal coordinates of the UAV at discrete time index $n$. $h[n]$ is the UAV altitude, and $N_{\mathrm{T}}$ is the total number of trajectory samples. Consecutive samples are separated by $\Delta t$ seconds.
% The mission consists of three phases. First, the UAV climbs from its initial altitude to the cruise altitude while remaining at the initial horizontal position. It then travels horizontally across multiple cellular coverage regions at a constant cruise altitude and horizontal speed $v_{\mathrm{UAV}}$. Finally, it descends at the destination while maintaining its final horizontal position. This construction separates altitude-dependent effects during climb and descent from horizontal variations encountered during cruise.

The horizontal (2D) and 3D distances between the UAV and sector $s$ at sample $n$ are given by:
\begin{align}
    d_{2\mathrm{D},s}[n]
    &=
    \sqrt{
        \left(x[n]-x_s\right)^2+
        \left(y[n]-y_s\right)^2
    },
    \label{eq:d2d}
    \\
    d_{3\mathrm{D},s}[n]
    &=
    \sqrt{
        d_{2\mathrm{D},s}^{2}[n]+
        \left(h[n]-h_{\mathrm{b}}\right)^2
    }.
    \label{eq:d3d}
\end{align}

The elevation and azimuth angles from sector $s$ toward the UAV are defined as follows:
\begin{align}
    \theta_s[n]
    &=
    \tan^{-1}
    \left(
        \frac{h[n]-h_{\mathrm{b}}}
        {d_{2\mathrm{D},s}[n]}
    \right),
    \label{eq:elevation}
    \\
    \phi_s[n]
    &=
    \operatorname{atan2}
    \left(
        y[n]-y_s,
        x[n]-x_s
    \right).
    \label{eq:azimuth}
\end{align}

\subsection{Sector Antenna Model}
\label{subsec:antenna}

Each sector employs a directional antenna characterized by
maximum gain $G_{\max}$, electrical downtilt
$\theta_{\mathrm{tilt}}$, horizontal 3-dB beamwidth
$\phi_{3\mathrm{dB}}$, and vertical 3-dB beamwidth
$\theta_{3\mathrm{dB}}$. The horizontal angular offset between the UAV direction $\phi_s[n]$ and the sector boresight direction $\psi_s$ is given by:
\begin{equation}
\Delta\phi_s[n]
=
\left(
\bigl(\phi_s[n]-\psi_s+180^{\circ}\bigr)
\bmod 360^{\circ}
\right)
-180^{\circ}.
\label{eq:horizontal_offset}
\end{equation}
This mapping produces the smallest signed angular separation
within $[-180^{\circ},180^{\circ}]$ and accounts for the
$0^{\circ}/360^{\circ}$ azimuth.

Under the convention that $\theta_{\mathrm{tilt}}>0$ denotes
downward electrical downtilt, the vertical angular offset is defined as:
\begin{equation}
\Delta\theta_s[n]
=
\theta_s[n]+\theta_{\mathrm{tilt}}.
\label{eq:vertical_offset}
\end{equation}

Let $A_m$ denote the maximum antenna attenuation and
$\mathrm{SLA}_V$ denote the vertical side-lobe attenuation
limit. Using the angular offsets in
\eqref{eq:horizontal_offset} and \eqref{eq:vertical_offset},
the horizontal and vertical attenuations are given by:
\begin{align}
A_{\mathrm{H},s}[n]
&=
\min\left\{
12\left(
\frac{\Delta\phi_s[n]}{\phi_{3\mathrm{dB}}}
\right)^2,
A_m
\right\},
\label{eq:horizontal_attenuation}\\
A_{\mathrm{V},s}[n]
&=
\min\left\{
12\left(
\frac{\Delta\theta_s[n]}{\theta_{3\mathrm{dB}}}
\right)^2,
\mathrm{SLA}_V
\right\}.
\label{eq:vertical_attenuation}
\end{align}
The composite attenuation and sector gain toward the UAV are then determined as follows:
\begin{align}
A_s[n]
&=
\min\left\{
A_{\mathrm{H},s}[n]+A_{\mathrm{V},s}[n],
A_m
\right\},
\label{eq:composite_attenuation}\\
% \end{equation}
%\begin{equation}
G_s[n]
&=
G_{\max}-A_s[n].
\label{eq:antenna_gain}
\end{align}

This antenna model accounts for the downtilted sector pattern
used in terrestrial cellular networks. Because the main lobe is
directed toward ground users, a UAV may be served through an
antenna sidelobe while also receiving interference from the
sidelobes of neighboring sectors.

\subsection{Urban Propagation Model}
\label{subsec:channel}

Large-scale propagation follows the 3GPP-based UMa LOS and NLOS models. For sector $s$ and trajectory sample $n$, the LOS probability is defined as follows:
\begin{equation}
    p_{\mathrm{L},s}[n]
    =
    P_{\mathrm{LOS}}
    \left(
        d_{2\mathrm{D},s}[n],
        h[n]
    \right),
\end{equation}
where the distance- and height-dependent LOS probability follows the aerial-user model adopted from \cite{3gpp38901,3gpp36777}.

The effective path loss is computed as:
\begin{equation}
L_s[n]
=
p_{\mathrm{L},s}[n]L_{\mathrm{LOS}}(\cdot)
+
\bigl(1-p_{\mathrm{L},s}[n]\bigr)L_{\mathrm{NLOS}}(\cdot),
\end{equation}
where $L_{\mathrm{LOS}}(\cdot)$ and $L_{\mathrm{NLOS}}(\cdot)$
denote the 3GPP LOS and NLOS path-loss functions~\cite{3gpp36777, 3gpp38901},
respectively.

% For each independent realization, the propagation state is generated as:
% \begin{equation}
%     Z_s[n]
%     =
%     \begin{cases}
%         \mathrm{LOS},
%         & U_s[n]\leq p_{\mathrm{L},s}[n],
%         \\
%         \mathrm{NLOS},
%         & U_s[n]>p_{\mathrm{L},s}[n],
%     \end{cases}
% \end{equation}
% % where $U_s[n]\sim\mathcal{U}(0,1)$. $\mathcal{U}(0,1)$ denotes a random variable uniformly distributed over the interval $[0,1]$. 
% The resulting path loss is given by:
% \begin{equation}
%     L_s[n]
%     =
%     \begin{cases}
%         L_{\mathrm{LOS}}
%         \left(d_{2\mathrm{D},s}[n],
%         d_{3\mathrm{D},s}[n],h[n]\right),
%         & Z_s[n]=\mathrm{LOS},
%         \\
%         L_{\mathrm{NLOS}}
%         \left(d_{2\mathrm{D},s}[n],
%         d_{3\mathrm{D},s}[n],h[n]\right),
%         & Z_s[n]=\mathrm{NLOS}, 
%     \end{cases}
% \end{equation}
% where $L_{\mathrm{LOS}}(\cdot)$ and $L_{\mathrm{NLOS}}(\cdot)$ denote the 3GPP LOS and NLOS path-loss functions~\cite{3gpp36777, 3gpp38901}, respectively.

Log-normal shadow fading (SF) is added to the probability-weighted
path loss as follows:
\begin{equation}
\widetilde{L}_s[n] = L_s[n] + X_s[n],
\qquad
X_s[n] \sim \mathcal{N}(0,\sigma_{\mathrm{SF}}^2),
\end{equation}
where $\sigma_{\mathrm{SF}}$ is the SF standard deviation listed in Table~I.
% Log-normal shadow fading is added to the path loss:
% \begin{equation}
%     \widetilde{L}_s[n]
%     =
%     L_s[n]+X_s[n],
%     \qquad
%     X_s[n]\sim\mathcal{N}(0,\sigma_{Z_s}^{2}),
% \end{equation}
% where $\sigma_{Z_s}$ depends on whether the link is LOS or NLOS.

Let $P_{\mathrm{tx}}$ denote the total transmit power of each
sector in dBm, and let $G_{\mathrm{UAV}}$ denote the UAV receive
antenna gain in dBi. The total received power from sector $s$
over the system bandwidth at trajectory sample $n$ is given by:
\begin{equation}
P_{\mathrm{rx},s}[n]
=
P_{\mathrm{tx}}
+
G_s[n]
+
G_{\mathrm{UAV}}
-
\widetilde{L}_s[n].
\label{eq:received_power}
\end{equation}

\section{Trajectory-Level Radio Serviceability}
\label{sec:serviceability}
This section formulates the trajectory-level radio metrics and serviceability measures. Received powers are converted to linear units when calculating interference, SINR, and RSRQ.

\subsection{Per-Sector RSRP Calculation}

The received power in~\eqref{eq:received_power} is converted from dBm to linear units in mW as follows:
\begin{equation}
p_{\mathrm{rx},s}[n]
=
10^{P_{\mathrm{rx},s}[n]/10}.
\label{eq:linear_received_power}
\end{equation}
Assuming that the sector transmit power is uniformly distributed across the $N_{\mathrm{RB}}$ resource blocks and their 12 subcarriers, the linear RSRP from sector $s$ is given by:
\begin{equation}
r_s[n]
=
\frac{p_{\mathrm{rx},s}[n]}{12N_{\mathrm{RB}}},
\label{eq:linear_rsrp}
\end{equation}
where $N_{\mathrm{RB}}$ is the number of resource blocks. The
corresponding RSRP in dBm is defined as:
\begin{equation}
R_s[n]
=
10\log_{10}\!\left(r_s[n]\right).
\label{eq:rsrp_dbm}
\end{equation}
These per-sector RSRP values are used for initial association
and handover decisions.

\subsection{Serving-Cell Association and Handover}

To preserve serving-cell continuity along the UAV trajectory,
an RSRP-based A3 handover procedure is applied. Let $c[n]$
denote the serving sector at sample $n$. The initial serving
sector is selected as:
\begin{equation}
c[0]=\arg\max_{s\in\mathcal{S}} R_s[0].
\end{equation}

At each subsequent sample, the sector providing the highest RSRP is identified as:
\begin{equation}
b[n]=\arg\max_{s\in\mathcal{S}} R_s[n].
\end{equation}

The A3 condition is satisfied when
\begin{equation}
b[n]\neq c[n-1],
\qquad
R_{b[n]}[n]>R_{c[n-1]}[n]+H_{\mathrm{A3}},
\end{equation}
where $H_{\mathrm{A3}}$ is the A3 handover offset. Let $N_{\mathrm{TTT}}=2$ denote the number of consecutive samples for which the same candidate must satisfy the A3 condition. With $\Delta t=0.1$~s, this corresponds to an effective time-to-trigger (TTT) of $0.2$~s.
% where $H_{\mathrm{A3}}$ is the A3 handover offset and
% $T_{\mathrm{TTT}}=160$~ms is the time-to-trigger (TTT), during which the same candidate sector must continuously satisfy the A3 condition before handover execution.
If the same candidate sector satisfies this condition, the serving sector is updated as:
\begin{equation}
c[n]=
\begin{cases}
b[n], & \text{if the A3 condition persists for }
%T_{\mathrm{TTT}},\\
N_{\mathrm{TTT}},\\
c[n-1], & \text{otherwise}.
\end{cases}
\end{equation}
The resulting serving-cell RSRP is given by:
\begin{equation}
R[n]=R_{c[n]}[n].
\end{equation}

\subsection{Interference, SINR, and RSRQ}
\label{subsec:sinr_rsrq}

Let $a_s[n]\in[0,1]$ denote the effective resource-overlap or activity factor of interfering sector $s$ at trajectory sample $n$. A value of $a_s[n]=1$ represents full co-channel activity, while a smaller value represents partial overlap between serving and interfering transmissions. The aggregate interference power is:
\begin{equation}
i[n] =
\sum_{\substack{s\in\mathcal{S}\\s\neq c[n]}}
a_s[n]p_{\mathrm{rx},s}[n].
\end{equation}

The receiver noise power over bandwidth $B$ is
\begin{equation}
    p_{\mathrm{N}}
    =
    10^{
    \left(
    -174+
    10\log_{10}B+
    F_{\mathrm{N}}
    \right)/10
    },
    \label{eq:noise_power}
\end{equation}
where $B$ is expressed in Hz, $-174$ dBm/Hz is the thermal-noise density, and $F_{\mathrm{N}}$ is the receiver noise figure in dB.

The serving-cell SINR in linear and dB scale is given by:
\begin{equation}
\gamma[n] =
\frac{p_{\mathrm{rx},c[n]}[n]}
{i[n]+p_{\mathrm{N}}},
\end{equation}
\begin{equation}
    \Gamma[n]
    =
    10\log_{10}
    \left(
        \gamma[n]
    \right).
    \label{eq:sinr_db}
\end{equation}

The received signal strength indicator (RSSI) accounts for the serving signal, aggregate interference, and receiver noise, and is given by:
\begin{equation}
p_{\mathrm{RSSI}}[n]
=
p_{\mathrm{rx},c[n]}[n]+i[n]+p_{\mathrm{N}}.
\end{equation}
RSRQ in linear and dB scale is then given by:
\begin{equation}
q[n] =
\frac{N_{\mathrm{RB}}r_{c[n]}[n]}
{p_{\mathrm{RSSI}}[n]},
\end{equation}
\begin{equation}
    Q[n]
    =
    10\log_{10}
    \left(
        q[n]
    \right).
    \label{eq:rsrq_db}
\end{equation}

Equations~\eqref{eq:rsrp_dbm}, \eqref{eq:sinr_db}, and \eqref{eq:rsrq_db} characterize the serving-cell RSRP,
SINR, and RSRQ, respectively, and distinguish received signal strength
from interference-limited signal quality.

\subsection{KPM-Specific and Joint Availability}
\label{subsec:availability}

Let $R_{\min}$, $Q_{\min}$, and $\Gamma_{\min}$ denote the selected evaluation thresholds for RSRP, RSRQ, and SINR, respectively. The corresponding binary service indicators at trajectory sample $n$ are given by:
\begin{align}
\mathcal{I}_{R}[n]
&=
\mathbf{1}\!\left\{R[n]\geq R_{\min}\right\},
\label{eq:rsrp_indicator}\\
\mathcal{I}_{Q}[n]
&=
\mathbf{1}\!\left\{Q[n]\geq Q_{\min}\right\},
\label{eq:rsrq_indicator}\\
\mathcal{I}_{\Gamma}[n]
&=
\mathbf{1}\!\left\{\Gamma[n]\geq\Gamma_{\min}\right\}.
\label{eq:sinr_indicator}
\end{align}
Here, $\mathbf{1}\{\cdot\}$ denotes the indicator function,
which equals one when the enclosed condition is satisfied and
zero otherwise.

The trajectory-level availability for KPM $k\in\{R,Q,\Gamma\}$ is given as follows:
\begin{equation}
    A_k
    =
    \frac{1}{N_{\mathrm{T}}}
    \sum_{n=0}^{N_{\mathrm{T}}-1}
    \mathcal{I}_k[n].
    \label{eq:kpi_availability}
\end{equation}
We define the joint radio service to be available only when all three KPM requirements are satisfied simultaneously, and we introduce the following metric to characterize this condition:
\begin{equation}
    \mathcal{I}_{\mathrm{joint}}[n]
    =
    \mathcal{I}_{R}[n]
    \mathcal{I}_{Q}[n]
    \mathcal{I}_{\Gamma}[n].
    \label{eq:joint_indicator}
\end{equation}

The joint service availability over the UAV mission is:
\begin{equation}
    A_{\mathrm{joint}}
    =
    \frac{1}{N_{\mathrm{T}}}
    \sum_{n=0}^{N_{\mathrm{T}}-1}
    \mathcal{I}_{\mathrm{joint}}[n].
    \label{eq:joint_availability}
\end{equation}

% Unless otherwise stated, the evaluation uses
% \begin{equation}
%     R_{\min}=-100~\mathrm{dBm},
%     \qquad
%     Q_{\min}=-12~\mathrm{dB},
%     \qquad
%     \Gamma_{\min}=0~\mathrm{dB}.
%     \label{eq:evaluation_thresholds}
% \end{equation}

% These values are used as evaluation thresholds for comparing deployment configurations. They are not presented as universal 3GPP service requirements. Because the numerical value of joint availability depends on the selected thresholds, Section~\ref{sec:results} also evaluates threshold sensitivity.
% \subsection{Coverage-Serviceability Gap}
% \label{subsec:coverage_gap}
Conventional coverage analysis commonly classifies a trajectory sample as covered when the RSRP threshold is satisfied. However, this classification does not ensure that RSRQ and SINR are also acceptable. We therefore define the coverage-serviceability gap as follows:
\begin{equation}
\Delta_{\mathrm{CS}}
=A_R-A_{\mathrm{joint}}.    
\end{equation}

% A small $\Delta_{\mathrm{CS}}$ indicates that RSRP coverage is a reasonable approximation of joint serviceability. A large $\Delta_{\mathrm{CS}}$ indicates that RSRP-only coverage substantially overestimates the portion of the UAV mission that meets both signal strength and signal quality requirements. The KPM that limits serviceability is defined as:
% \begin{equation}
%     k^{\star}
%     =
%     \arg\min_{k\in\{R,Q,\Gamma\}}
%     A_k.
%     \label{eq:serviceability_bottleneck}
% \end{equation}

% This bottleneck formulation distinguishes whether service is primarily constrained by insufficient received power, poor reference-signal quality, or low SINR.

\section{Numerical Results}
\label{sec:results}
This section evaluates the difference between conventional RSRP coverage and joint radio serviceability along the urban UAV trajectory. We first examine how RSRP and RSRQ evolve with altitude, then analyze the SINR distribution and the effect of ISD. Finally, we quantify KPM-specific availability, joint availability, and the coverage-serviceability gap.

\subsection{Evaluation Configuration}
\label{subsec:evaluation_setup}
The simulation parameters are summarized in Table~\ref{tab:simulation_params}. The network contains 19 sites and 57 co-channel sectors operating at 3.5~GHz with a bandwidth of 20~MHz. The UAV follows the climb-cruise-descent trajectory described in Section~\ref{subsec:trajectory}, with altitudes ranging from 5~m to 300~m. The analysis considers ISDs of 500 and 1000~m.
\begin{table}[t]
\centering
\caption{Urban 5G NR simulation parameters.}
\label{tab:simulation_params}
\renewcommand{\arraystretch}{1.08}
\setlength{\tabcolsep}{3.5pt}
\footnotesize
\begin{tabular}{cc|cc}
\hline
\textbf{Parameter} & \textbf{Value}
& \textbf{Parameter} & \textbf{Value} \\
\hline
$f_c$                    & 3.5 GHz
& $B$                    & 20 MHz \\
$N_{\mathrm{RB}}$        & 51
& $P_{\mathrm{tx}}$      & 46 dBm \\
$h_{\mathrm{b}}$         & 25 m
& $G_{\max}$             & 17 dBi \\
$\phi_{3\mathrm{dB}}$    & $65^{\circ}$
& $\theta_{3\mathrm{dB}}$ & $30^{\circ}$ \\
$\theta_{\mathrm{tilt}}$ & $6^{\circ}$
& $A_{\mathrm{m}}$       & 15 dB \\
$\mathrm{SLA}_{\mathrm{V}}$ & 15 dB
& $G_{\mathrm{UAV}}$     & 3 dBi \\
$F_{\mathrm{N}}$         & 7 dB
& $D_{\mathrm{ISD}}$     & 500, 1000 m \\
$h[n]$                   & 5--300 m
& $v_{\mathrm{UAV}}$     & 5 m/s \\
$\Delta t$               & 0.1 s
%& $T_{\mathrm{TTT}}$     & 160 ms \\
&$N_{\mathrm{TTT}}$ & 2 samples\\
$R_{\min}$               & $-100$ dBm
& $Q_{\min}$             & $-18$, $-22$ dB \\
$\Gamma_{\min}$          & $0$, $-6$ dB
& $H_{\mathrm{A3}}$      & 3 dB \\
$\sigma_{\mathrm{SF}}$  & 6 dB 
&$a_s[n]$ & $0.5$\\
\hline
\end{tabular}
\vspace{-.2in}
\end{table}
The strongest-RSRP sector is used for initial association. During the
remainder of the trajectory, the current serving sector is retained
until another sector satisfies the A3 handover condition. 
%The resulting serving-sector sequence is used to calculate serving-cell RSRP, RSRQ, and SINR. Guided by practical LTE/5G signal-quality categories reported in~\cite{shakir2023kpi,Polak24082538}, we use $-100$~dBm for RSRP as a cell-edge reference, and $-20$~dB for RSRQ and $0$~dB for SINR as poor-quality boundaries. 
For the service-availability analysis, we consider an RSRP threshold of $-100$~dBm and two evaluation configurations with RSRQ and SINR thresholds of $(-18, 0)$~dB and $(-22, -6)$~dB, respectively. The same underlying radio measurements are used for both configurations to examine the sensitivity of joint service availability to the selected signal-quality requirements. For each ISD, three independent simulation realizations with different random seeds are evaluated. The reported altitude-dependent KPM values are first calculated for each realization and then averaged across realizations, while the service-availability metrics are computed separately for each realization and averaged.

\subsection{Altitude-Dependent Coverage-Quality Decoupling}
\label{subsec:coverage_quality_results}
\begin{figure}[!t]
    \centering
    \includegraphics[width=\columnwidth,trim={0in 0.15in 0in 0in},clip]
    %{figs/new/rsrp_rsrq_vs_altitude_urban2.pdf}
    {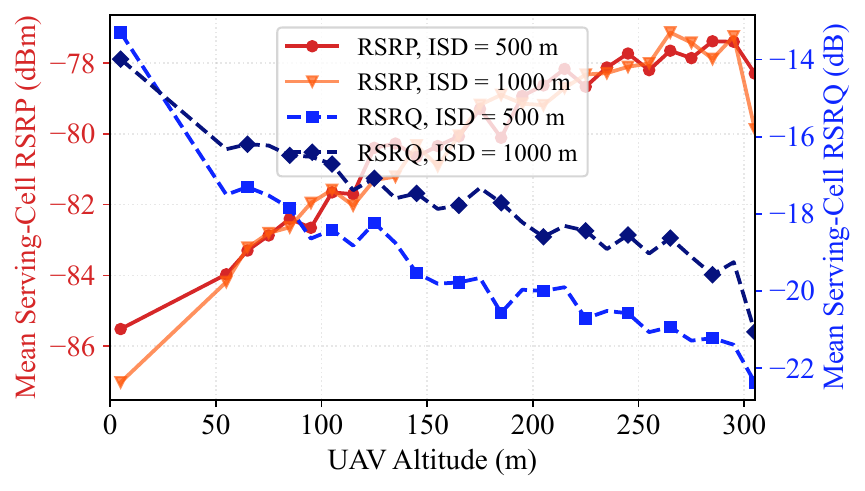}
    \caption{Mean serving-cell RSRP and RSRQ versus UAV altitude for ISDs of 500 and 1000~m, averaged over three independent simulation realizations. Solid lines denote RSRP, while dashed lines denote RSRQ.}
    \label{fig:rsrp_rsrq_altitude}
    \vspace{-.3in}
\end{figure}

Fig.~\ref{fig:rsrp_rsrq_altitude} compares the mean serving-cell RSRP and RSRQ as functions of UAV altitude, where each point represents the average across the independent simulation realizations. In the urban environment, RSRP exhibits an overall increasing trend with altitude. At low altitude, buildings and surrounding clutter increase attenuation and reduce the probability of a strong serving link. As altitude increases, reduced blockage and greater LOS visibility strengthen the received serving-cell signal. %Across the evaluated trajectory, RSRP increases from approximately $-92$~dBm at low altitude to values between approximately $-78$ and $-74$~dBm near 300~m.
Across the evaluated trajectory, as shown in Fig.~\ref{fig:rsrp_rsrq_altitude}, at low altitude, mean RSRP is approximately $-85.5$ dBm and $-87$ dBm for ISDs of $500$ m and $1000$ m, respectively, and improves to approximately $-78$ dBm at higher altitudes. In contrast, at low altitude, mean RSRQ is approximately $-13.5$ dB for ISD = 500~m and $-14$ dB for ISD = 1000 m, decreasing to about $-22$ dB and $-21$ dB, respectively, at higher altitudes.
% mean RSRQ decreases from approximately $-13.5$ to $-14$ dB at low altitude to about $-22$ dB and $-21$ dB near the upper altitude range for ISDs of $500$ m and $1000$ m, respectively.
%The corresponding RSRQ behavior is markedly different. RSRQ is approximately $-12$ to $-13$~dB at low altitudes but decreases toward approximately $-18$ to $-20$~dB at higher altitudes. 
Hence, the UAV receives a stronger serving signal as it ascends, but it becomes visible to a larger number of interfering sectors. Aggregate interference grows relative to the desired reference signal, degrading both RSRQ and SINR.

The ISD comparison further exposes the coverage-quality trade-off. The 500 m deployment generally provides stronger RSRP because the serving BSs are closer to the UAV. In contrast, the 1000 m deployment generally provides better RSRQ at medium and high altitudes because greater BS separation reduces aggregate ICI. Thus, network densification can improve aerial received power without improving, and potentially while degrading, aerial signal quality. These results establish a key observation of the study: increasing altitude can improve nominal urban UAV coverage while simultaneously reducing interference-sensitive radio quality. Consequently, altitude-dependent RSRP alone does not reliably indicate whether the UAV receives usable cellular service.

\subsection{SINR Degradation with Altitude and Effect of ISD}
\begin{figure}[!t]
    \centering    %\includegraphics[width=1\columnwidth]    %{figs/new/urban_sinr_vs_altitude2.pdf}
    \includegraphics[width=\columnwidth,trim={0in 0.1in 0in 0in},clip]{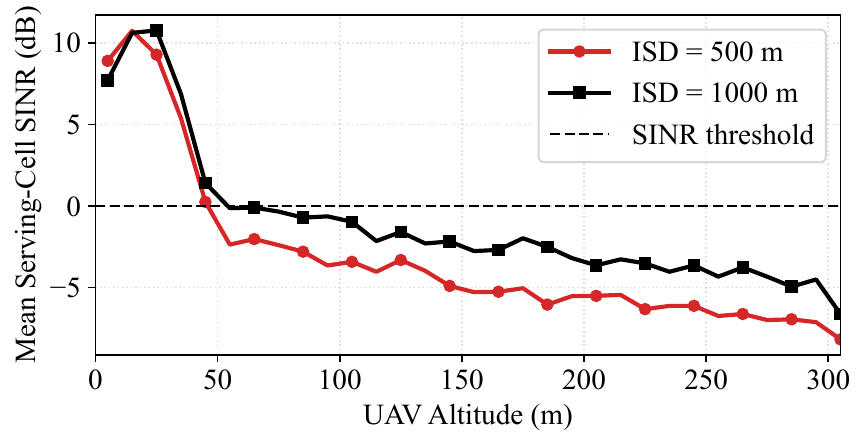}
    \caption{Mean serving-cell SINR versus UAV altitude for ISDs of 500 and 1000~m, averaged over three independent simulation realizations. The horizontal dashed line represents the $0$-dB SINR threshold used in Fig.~\ref{fig:kpi_joint_availability_rsrq_18dB_SINR_0dB}.}
    \vspace{-.2in}
    \label{fig:sinr_altitude}
\end{figure}

Fig.~\ref{fig:sinr_altitude} presents the mean serving-cell SINR as a function of UAV altitude across the independent simulation realizations. For both deployments (ISD = 500 m and 1000 m), SINR generally decreases as altitude increases. As altitude increases, improved LOS visibility exposes the UAV to additional interfering sectors, causing SINR to decrease despite stronger serving-cell RSRP. At higher altitudes, SINR remains predominantly below the $0$-dB reference threshold.
%At low altitudes, surrounding clutter limits the visibility of distant co-channel sectors. As the UAV ascends, improved LOS visibility strengthens the serving signal but also exposes the UAV to a larger number of interfering sectors. The increase in aggregate interference dominates the serving-signal improvement, causing SINR to decrease and remain predominantly below the selected $0-$dB \textcolor{blue}{reference} threshold at higher altitudes.
The 1000-m deployment provides higher SINR than the 500-m deployment over most of the evaluated altitude range because greater separation between co-channel sites reduces aggregate interference. Nevertheless, SINR remains below the selected threshold over most of the trajectory for both deployments, demonstrating that strong serving-cell RSRP does not necessarily imply an interference-tolerant aerial link.

% The 1000-m deployment provides higher SINR than the 500-m deployment over most of the evaluated altitude range because greater separation between co-channel sites reduces aggregate interference. This improvement is consistent with
% the greater separation between co-channel sites, which reduces
% the aggregate interference observed by the UAV. Nevertheless,
% SINR remains unavailable over most of the trajectory for both
% deployments, demonstrating that strong serving-cell RSRP does
% not necessarily imply an interference-tolerant aerial link.

\subsection{KPM-Specific and Joint Service Availability}
\label{subsec:availability_results}
% \begin{figure}[!t]
%     \centering
%     \includegraphics[width=\columnwidth]
%     %{figs/new/urban_kpi_joint_availability_vs_isd2.pdf}
%     {figs/new/urban_kpi_joint_availability_multiple_realizations2.pdf}
%     \caption{Mean KPM-specific and joint service availability across three independent realizations for different ISDs.}
%     \label{fig:kpi_joint_availability}
%     \vspace{-.1in}
% \end{figure}
\begin{figure}[!t]
    \centering
    \includegraphics[width=\columnwidth,trim={0in 0.1in 0in 0in},clip]{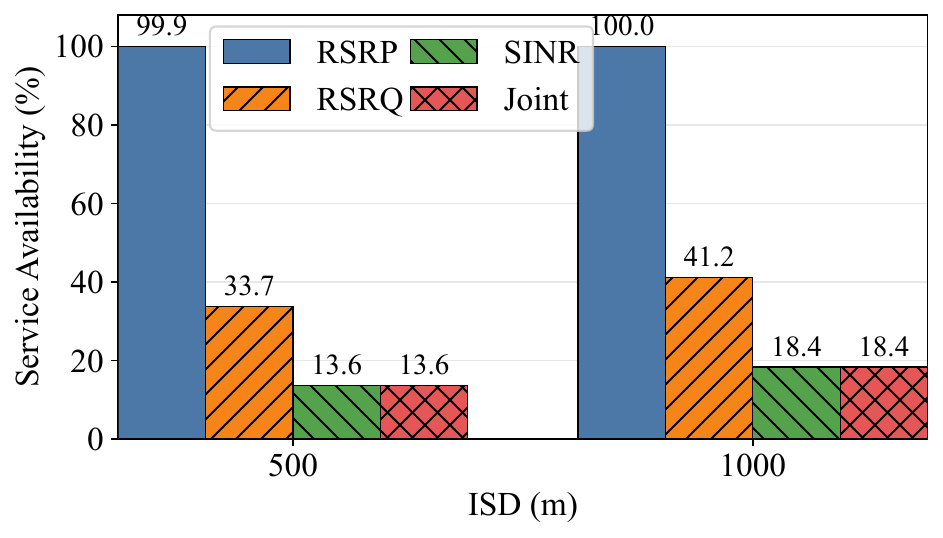}
    \caption{Joint serviceability under an RSRQ threshold of $-18$ dB and an SINR threshold of $0$ dB.}
    \label{fig:kpi_joint_availability_rsrq_18dB_SINR_0dB}
    \vspace{-.2in}
\end{figure}
\begin{figure}[!t]
    \centering
    \includegraphics[width=\columnwidth,trim={0in 0.1in 0in 0in},clip]    {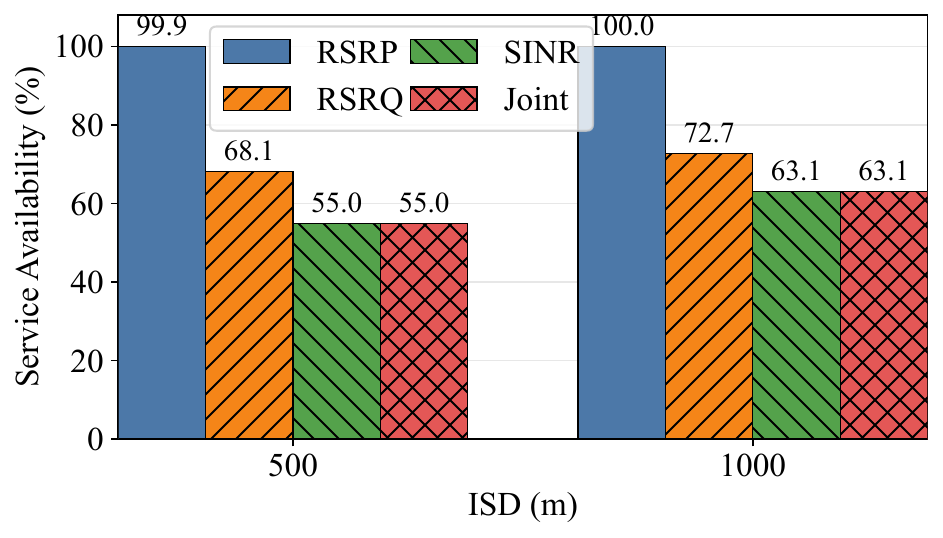}
    \caption{Joint serviceability under an RSRQ threshold of $-22$ dB and an SINR threshold of $-6$ dB.}    \label{fig:kpi_joint_availability_rsrq_22dB_SINR_6dB}
    \vspace{-.3in}
\end{figure}
Figs.~\ref{fig:kpi_joint_availability_rsrq_18dB_SINR_0dB} and
\ref{fig:kpi_joint_availability_rsrq_22dB_SINR_6dB} compare the KPM-specific
and joint service availability under two evaluation configurations. The RSRP threshold is fixed at $-100$~dBm, while the RSRQ and SINR thresholds are set to $(-18, 0)$~dB and $(-22, -6)$~dB, respectively. As shown in Fig.~\ref{fig:kpi_joint_availability_rsrq_18dB_SINR_0dB}, RSRP availability remains nearly $100\%$ for both ISDs. However, at ISDs of $500$~m and $1000$~m, RSRQ availability is $33.7\%$ and $41.2\%$, respectively, while SINR availability is limited to $13.6\%$ and $18.4\%$. Joint service availability consequently equals SINR availability, yielding coverage-serviceability gaps of $86.3$ and $81.6$ percentage points, respectively. Under the second evaluation configuration, Fig.~\ref{fig:kpi_joint_availability_rsrq_22dB_SINR_6dB} shows that RSRQ availability increases to $68.1\%$ and $72.7\%$ for ISDs of $500$~m and $1000$~m, respectively. The corresponding SINR and joint availability increase to $55.0\%$ and $63.1\%$. The resulting coverage-serviceability gaps are $44.9$ and $36.9$ percentage points, respectively. These results demonstrate that joint service availability depends on the selected signal-quality requirements. Although the second configuration provides higher joint availability, nearly complete RSRP coverage does not translate into equivalent
joint service availability under either configuration. SINR remains the limiting metric for both ISDs, highlighting the importance of multi-KPM evaluation for UAV radio serviceability.
%Fig.~\ref{fig:kpi_joint_availability} compares RSRP, RSRQ, SINR, and joint service availability. For an ISD of 500~m, the RSRP requirement is satisfied during 99.9\% of the UAV trajectory. An RSRP-only assessment would therefore classify almost the entire mission as covered. However, SINR availability is only 13.6\%, and RSRQ availability is only 52.7\%. Because joint service requires all three conditions to be satisfied simultaneously, joint availability is also 13.6\%. The resulting coverage-serviceability gap is $\Delta_{\mathrm{CS}}=86.3$ percentage points. This large gap demonstrates that nominal RSRP coverage substantially overestimates the portion of the UAV mission satisfying both signal-strength and signal-quality requirements. 
% Increasing the ISD from 500~m to 1000~m improves mean SINR availability from 13.6\% to 18.4\% and RSRQ availability from 52.7\% to 61.2\%. Mean joint availability correspondingly increases from 13.6\% to 18.4\%, as reduced aggregate interference improves SINR, but most trajectory samples still fail the SINR requirement. The KPM-specific comparison is important because it identifies the origin of service failure. RSRP is not the limiting factor in either deployment. Instead, aerial serviceability is constrained by interference-sensitive measurements, particularly SINR. This observation indicates that extending RSRP coverage or increasing serving-cell power alone would not resolve the serviceability limitation.

% Add the 1500 m and 2000 m availability values here.
% Report the coverage-serviceability gap separately for every ISD.

\section{Conclusion}
\label{sec:conclusion}
This paper presented a trajectory-aware, 3GPP-based system-level analysis of cellular-connected UAV serviceability in urban 5G NR networks. By jointly evaluating RSRP, RSRQ, and SINR, we quantified the coverage-serviceability gap along a 3D trajectory. Under RSRQ and SINR thresholds of $-18$~dB and $0$~dB, respectively, joint availability was limited to $13.6\%$ and $18.4\%$ for ISDs of $500$~m and $1000$~m, despite nearly complete RSRP coverage. With thresholds of $-22$~dB and $-6$~dB, joint availability increased to $55.0\%$ and $63.1\%$, respectively. These results demonstrate that RSRP--only coverage substantially overestimates mission-level UAV serviceability under both evaluation configurations, with SINR remaining the limiting metric. The findings motivate trajectory-aware, multi-KPM aerial radio planning. Future work will consider maneuver-rich UAV missions to assess their impact on radio-link quality and handover behavior.
%At an ISD of 500~m, 99.9\% mean RSRP availability translated into only 13.6\% mean joint availability, yielding an 86.3-percentage-point gap. Increasing the ISD to 1000~m improved joint availability to 18.4\% by reducing aggregate interference, although SINR remained the dominant bottleneck. These results show that RSRP-only coverage can substantially overestimate mission-level UAV connectivity, motivating trajectory-aware, multi-KPM aerial radio planning.

%Future work will consider maneuver-rich UAV missions incorporating abrupt stop-and-go behavior, direction changes, variable speeds, and trajectory deviations to assess their impact on radio-link quality and handover behavior.

% In this paper, we present a 3GPP-based system-level analysis of altitude-dependent performance for cellular-connected UAVs in 5G NR networks, incorporating realistic deployment geometry, antenna patterns, and probabilistic LOS/NLOS propagation. Our results show a clear transition from coverage-limited behavior at low altitudes to interference-limited behavior at higher altitudes. While serving-cell RSRP generally improves with altitude because of
% reduced blockage and increased LOS probability, RSRQ and SINR can
% deteriorate because the UAV becomes visible to a larger number of
% co-channel sectors. Increasing the ISD weakens the received signal
% but can improve interference-sensitive KPMs by separating the
% interfering sites.
\bibliographystyle{IEEEtran}
\bibliography{references}

@techreport{3GPP36777,
  author       = {{3GPP TR 36.777}},
  title        = {Study on Enhanced {LTE} Support for Aerial Vehicles},
  institution  = {3GPP},
  number       = {TR 36.777},
  year         = {2017}
}

@techreport{3gpp38901,
  author={{3GPP TR 38.901}},
  title={{Study on channel model for frequencies from 0.5 to 100 GHz}},
  institution={3GPP},
  number={TR 38.901},
  year={2022}
}

@ARTICLE{Maeng11003933,
  author={Maeng, Sung Joon and Guvenc, Ismail},
  journal={IEEE Internet Things J.}, 
  title={Altitude-Dependent Cellular Spectrum Occupancy: From Measurements to Stochastic Geometry Models}, 
  year={2025},
  volume={12},
  number={15},
  pages={30268-30281},
  doi={10.1109/JIOT.2025.3570266}}

@ARTICLE{Jie10587227,
  author = {Seah, Shi Jie and Leow, Chee Yen and Nalinggam, Renuka and
          New, Wee Kiat and Alam, Rozana and Jong, Siat Ling and
          Lam, Hong Yin and Almasoud, Abdullah M.},  
  journal={IEEE Access}, 
  title={Empirical Channel Models for {UAV} Communication: A Comparative Study}, 
  year={2024},
  volume={12},
  number={},
  pages={96740-96756},
  doi={10.1109/ACCESS.2024.3424544}}

@article{mozaffari2019tutorial,
  title={A tutorial on {UAVs} for wireless networks: Applications, challenges, and open problems},
  author={Mozaffari, Mehdi and Saad, Walid and Bennis, Mehdi and Nam, Yong-Hoon and Debbah, Merouane},
  journal={IEEE Commun. Surveys Tuts.},
  volume={21},
  number={3},
  pages={2334--2360},
  year={2019}
}

@article{fotouhi2019survey,
  title={Survey on {UAV} cellular communications: Practical aspects, standardization advancements, regulation, and security challenges},
  author={Azade Fotouhi and
                  Haoran Qiang and
                  Ming Ding and
                  Mahbub Hassan and
                  Lorenzo Galati{-}Giordano and
                  Adrian Garc{\'{\i}}a{-}Rodr{\'{\i}}guez and
                  Jinhong Yuan},
  journal={IEEE Commun. Surveys Tuts.},
  volume={21},
  number={4},
  pages={3417--3442},
  year={2019}
}

@article{Chowdhury2021ReliableUAV,
  author  = {Md Moin Uddin Chowdhury and Ismail Guvenc and Walid Saad and Arupjyoti Bhuyan},
  title   = {Ensuring Reliable Connectivity to Cellular-Connected {UAVs} with Up-Tilted Antennas and Interference Coordination},
  journal = {ITU J. Future Evol. Technol.},
  volume  = {2},
  number  = {2},
  pages   = {165--185},
  year    = {2021},
  doi     = {10.52953/LEFT7402}
}

@ARTICLE{wang2022bscoop,
  author={Wang, Zhe and Zheng, Jun},
  journal={IEEE Trans. Veh. Technol.}, 
  title={Performance Modeling and Analysis of Base Station Cooperation for Cellular-Connected {UAV} Networks}, 
  year={2022},
  volume={71},
  number={2},
  pages={1807-1819},
  doi={10.1109/TVT.2021.3123826}}

@INPROCEEDINGS{hossen_booster-cell2026,
    author ={M. S. Hossen and V. K. Shah and I. Guvenc},
    title = {Aerial Booster-Cell Enabled Inter-Cell Interference Coordination for {5G NR} Networks},
    booktitle={Proc. IEEE Veh. Technol. Conf. (VTC)}, 
    year = {Sep. 6--9, 2026},
    pages = {1-7},
    address = {Boston, MA, USA},    
}

\end{document}